\documentclass{PoS}

\usepackage{psfrag}

\title{Exploring the infrared Landau gauge propagators using large asymmetric lattices}

\ShortTitle{Exploring the infrared Landau gauge propagators using large asymmetric lattices}

\author{\speaker{P. J. Silva} and O. Oliveira\\
        Centro de F\'{\i}sica Computacional,
        Departamento de F\'{\i}sica,
        Universidade de Coimbra, P-3004-516 Coimbra,
        Portugal\\
        E-mail: \email{psilva@teor.fis.uc.pt}, 
        \email{orlando@teor.fis.uc.pt}}

\abstract{We report on the infrared limit of the
quenched lattice Landau gauge gluon and ghost propagators computed from large
asymmetric lattices. In particular, the compatibility of the pure power
law infrared solutions of the Dyson-Schwinger equations with the lattice data
is investigated and the exponent $\kappa$ is measured. The gluon lattice
data favour $\kappa \sim 0.52$, which would imply a
vanishing zero momentum gluon propagator. For the subset of lattices where
the ghost propagator was computed, the data are not compatible with a pure
power law. Our data also show a decreasing running coupling in the
infrared region. Furthermore, positivity violation for the gluon
propagator is also verified.

}

\FullConference{XXIVth International Symposium on Lattice Field Theory\\
                July 23-28, 2006\\
                Tucson, Arizona, USA}

\begin{document}

\section{Introduction and motivation}

The study of the infrared limit of Quantum Chromodynamics (QCD) 
requires the use of 
non-perturbative methods. Two first principles non-perturbative 
approaches are Dyson-Schwinger equations (DSE) and the lattice 
formulation of QCD. Because both methods have good and bad features, 
a comparison between their results is a good test of our understanding 
of the low energy limit of QCD. During the last years, there has been an effort to compute the infrared 
gluon and ghost propagators in Landau gauge,

\begin{eqnarray}
  D^{ab}_{\mu\nu} (q) & = & \delta^{ab}  \,
      \left( \delta_{\mu\nu} - \frac{q_\mu q_\nu}{q^2} \right) \, 
  D (q^2) \, , \\
  G^{ab}(q) & = & - \, \delta^{ab}  \,
                                  G (q^2) \, ,
\end{eqnarray}

\noindent
and a running coupling constant\footnote{See \cite{Prosperi06} for a recent 
review on the running coupling constant.} defined from these propagators

\begin{equation}
  \alpha_S ( q^2 ) ~ = ~ \alpha_S ( \mu^2 ) \, Z^2_{ghost} ( q^2 ) \,
                                               Z_{gluon} ( q^2 ) \, ;
\label{alfaS}
\end{equation}

\noindent
$Z_{ghost} ( q^2 ) = q^2 G( q^2 )$ and 
$Z_{gluon} ( q^2 ) = q^2 D( q^2 )$ are the ghost and gluon dressing functions.

In \cite{lerche}, assuming ghost dominance, it was computed a solution of the DSE that predicts pure power laws for the propagators, namely

\begin{equation}
Z_{gluon}(q^2)\sim(q^2)^{2\kappa}\,,\,Z_{ghost}(q^2)\sim(q^2)^{-\kappa},
\end{equation}

\noindent
with $\kappa=0.595$. This implies a vanishing (infinite) gluon 
(ghost) propagator for zero momentum. 
Other studies of the infrared limit also predict 
$\kappa>0.5$ \cite{pawl,gies,tisq}.

\begin{figure}[!t]
\begin{center}
\vspace*{0.7cm}
\psfrag{EIXOX}{{\small $q(GeV)$}}
\psfrag{EIXOY}{{\small $q^2 D(q^2)$}}
\includegraphics[width=8cm]{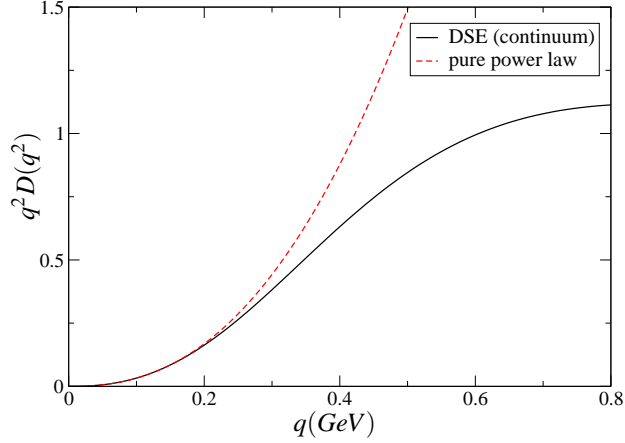}
\caption{The gluon DSE solution \cite{fish.penn} compared 
with the corresponding pure power law.}
\label{dsepower}
\end{center}
\end{figure}

As an infrared analytical solution of the DSE, the pure power laws 
are valid only 
for very low momenta. Indeed, comparing the DSE solution 
for the gluon propagator 
\cite{fish.penn} with the corresponding pure power law, 
see figure \ref{dsepower}, it comes out that the 
power law is valid only for momenta below $200$ MeV.
 
For lattice QCD, it is a challenge to perform a simulation with a 
minimum number of points in the region of interest. The symmetric 
lattices available at the moment have a limited number of points 
in the infrared region (see, for example, \cite{phdstern,berlinrio,sterntucson}).

In the series of papers \cite{sardenha, dublin, finvol, nossoprd, 
madrid, brasil} we have been using large asymmetric lattices 
$L_s^3 \times L_t$, with $L_t \gg L_s$, to investigate the infrared limit of the gluon and ghost propagators.
In this article we report on 
the status of our investigations concerning the use of asymmetric lattices 
to study the infrared properties of QCD.

\section{Gluon propagator}

In \cite{nossoprd}, we have computed the gluon propagator 
for SU(3) four-dimensional asymmetric lattices 
$L_s^3 \times 256$, with $L_s=8,10,\ldots,18$. 

As reported in \cite{finvol, nossoprd, madrid, brasil, cucc06, phdstern}, 
there are clear finite volume effects. However, 
the various simulations performed up to now show that the approach 
to the infinite volume limit $L_{\hspace{0.5mm}s} \to +\infty$ is smooth.

\begin{figure}[!t]
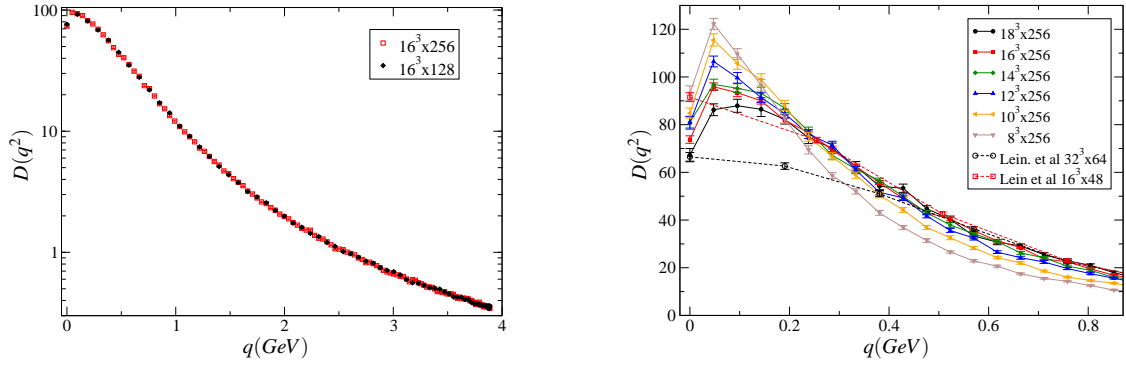

\psfrag{EIXOX}{{\scriptsize $q(GeV)$}}
\psfrag{EIXOY}{{\scriptsize $D(q^2)$}}
   \begin{minipage}[b]{0.45\textwidth}
   \centering
   \includegraphics[width=6.5cm]{raw_Ls_16.eps}
  \end{minipage} 
\hspace{0.08\textwidth}
  \begin{minipage}[b]{0.45\textwidth}
  \centering
  \includegraphics[width=6.5cm]{prop_all.V_Lein.eps}
  \end{minipage} 
\caption{On the left, the gluon propagator for $16^3\times 128$ 
and $16^3\times 256$ lattices, considering 
only pure temporal momenta. Note the logarithmic scale 
in the vertical axis. On the right, the gluon propagator for 
all lattices $L_{s}^{3}\times256$. For comparisation, we also show the 
$16^3\times 48$ and $32^3\times 64$ propagators computed in \cite{lein99}. } 
\label{asygluon}
\end{figure}

In what concerns the finite volume effects, our simulations suggest that 
the finite volume effects are essentially due to a relatively small 
spatial extension. Indeed, the gluon propagator was computed for 
$16^3\times128$ 
and $16^3\times256$, and the data are undistinguishable --- see figure 
\ref{asygluon} (left). This result gives us confidence that the 
temporal size of our lattices is sufficiently large.

However, it was observed that the propagator 
depends on the spatial size of the lattice 
--- see figure \ref{asygluon} (right). 
The gluon propagator decreases with the volume for the smallest momenta 
and increases with the volume for higher momenta.

\subsection{Infrared exponent}

In order to compute the infrared exponent $\kappa$ from the 
lattice data, we considered fits of the smallest temporal momenta 
of the gluon dressing function $Z_{gluon}(q^2)=q^2 D(q^2)$ to 
a pure power law with and without polinomial corrections\footnote{The results 
can be seen in table II of \cite{nossoprd}.}. In general, the $\kappa$ 
values increase with the volume 
of the lattice. So, our $\kappa$ can be read as lower bounds in the 
infinite volume figure $\kappa_{\hspace{0.5mm}\infty}$.

The gluon propagator can be extrapolated to $L_s \to +\infty$, 
as a function of the inverse of the volume,
fitting each timelike momentum propagator 
separately, and assuming a sufficient number of points 
in the temporal direction. Several types of polinomial extrapolations were 
tried, using different sets of lattices, and we conclude that the data are 
better described by quadratic extrapolations of the data from the 
4th and 5th largest lattices. 

The values of $\kappa$ extracted from these extrapolated propagators are 
$\kappa = 0.5215(29)$, with a $\chi^2/d.o.f. = 0.02$, using the largest 
5 lattices in the extrapolation, and $\kappa = 0.4979(66)$, 
$\chi^2 / d.o.f. = 0.27$ using the largest 4 lattices. Fitting the 
extrapolated data to the polinomial corrections to the pure power law, 
one gets higher values for $\kappa$. The first value $\kappa = 0.5215(29)$ 
is on 
the top of the value obtained from extrapolating directly $\kappa$ as a 
function of the volume \cite{nossoprd}. Note also that in 
\cite{dublin,brasil}, we fitted the gluon temporal data for larger 
ranges of momenta, using other model functions, giving always values 
for $\kappa$ above 0.5, supporting again an infrared vanishing gluon 
propagator. In conclusion, one can claim a $\kappa \in [0.49,0.53]$, 
with the lattice data favouring the right hand side of the interval.

\subsection{Positivity violation}

In QCD, the violation of reflection positivity for the Landau gauge gluon propagator means that the gluon cannot appear as a free asymptotic S-matrix state. This may be viewed as an indication of gluon confinement. 

On the lattice, one can study positivity violation for the gluon propagator from the real space propagator,
\begin{equation}
  C(t) ~ \sim ~ \int_{-\infty}^{\infty} dp D(p,\vec{p}=0) \exp(-ipt).
\end{equation}
Finding $C(t)<0$ for some $t$ is a sign of positivity violation.

\begin{figure}[!t]
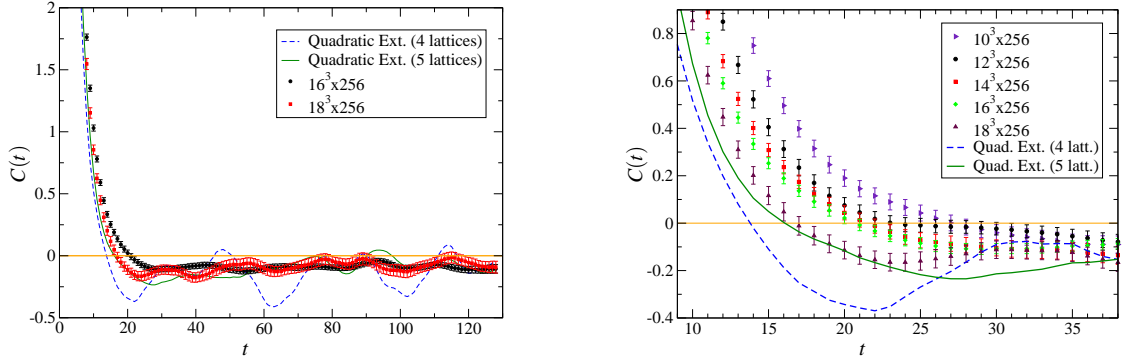

\psfrag{EIXOX}{{\scriptsize $t$}}
\psfrag{EIXOY}{{\scriptsize $C(t)$}}
   \begin{minipage}[b]{0.45\textwidth}
   \centering
   \includegraphics[width=6.5cm]{positivity_extrapol.eps}
  \end{minipage} 
\hspace{0.08\textwidth}
  \begin{minipage}[b]{0.45\textwidth}
  \centering
  \includegraphics[width=6.5cm]{positivity.zoom.256.eps}
  \end{minipage} 
\caption{On the left, the real space gluon propagator $C(t)$ for our 
largest lattices and for the two extrapolations considered in 
\cite{nossoprd}. On the right, a zoom of the region of interest 
showing all lattices. \label{posvio}} 
\end{figure}

Positivity violation for the gluon propagator has been observed for the
Dyson-Schwinger gluon propagator \cite{fischer04}, as well as for 
the propagator computed from symmetric lattices \cite {cuccposv,phdstern}. 
Here we show the real space propagator computed from our
asymmetric lattices, including in figure \ref{posvio}, $C(t)$ for 
the two infinite volume extrapolations of the gluon propagator 
computed in \cite{nossoprd}.

We thereby confirm that positivity violation occurs for the gluon 
propagator. Our data show that the time for positivity violation to
happen, decreases
when the spatial lattice volume is increased. Furthermore, the 
infinite volume limit suggests that positivity violation shows 
up at $t\sim1.5fm$. Previous studies show similar values 
\cite{phdstern, fischer04}. Moreover, similarly to what 
was observed in \cite{cuccposv}, for large time separations our data show an 
oscilatory behaviour.

\section{Ghost propagator}

So far, we have computed the ghost propagator for our smallest lattices \cite{madrid} ($10^3\times256$, $12^3\times256$, $16^3\times128$), 
using both a point source 
method \cite{suman} (we averaged over 7 different 
point sources to get a better 
statistics), and a plane-wave source \cite{cucc97}. In both cases we used 
the pre-conditioned conjugate gradient algorithm, 
as described in \cite{stern05}.
The plane-wave source method provides better statistical accuracy, but we can only obtain one momentum 
component at a time. The point source method allows to get all the momenta 
in one go, but with larger statistical errors.

\begin{figure}[!h]
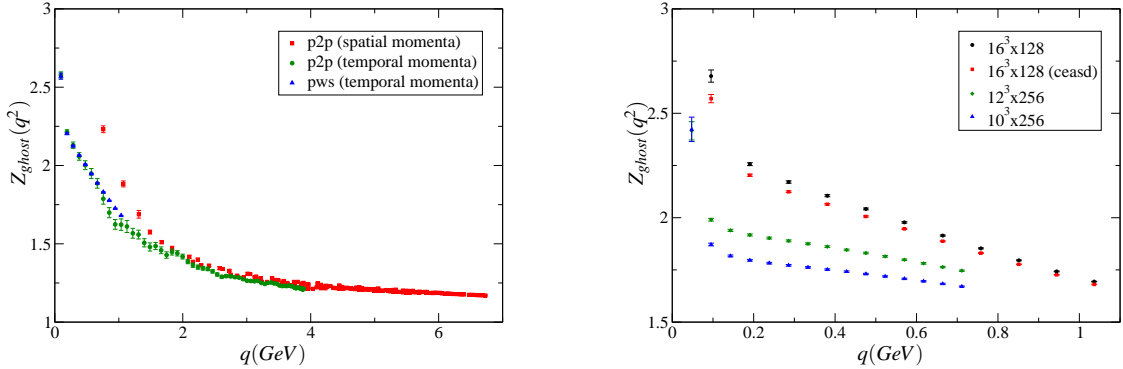

\psfrag{EIXOX}{{\scriptsize $q(GeV)$}}
\psfrag{EIXOY}{{\scriptsize $Z_{ghost}(q^2)$}}
   \begin{minipage}[b]{0.45\textwidth}
   \centering
   \includegraphics[width=6.5cm]{dress.16.3.128.ceasd.eps}
  \end{minipage} 
\hspace{0.08\textwidth}
  \begin{minipage}[b]{0.45\textwidth}
  \centering
  \includegraphics[width=6.5cm]{cucc.eps}
  \end{minipage} 
\caption{ On the left, the bare ghost dressing function for the 
$16^3\times128$ lattice, gauge fixed with CEASD method. ``p2p'' (``pws'') 
stands for the ghost components computed using a 
point (plane wave) source. On the right, the bare ghost dressing 
function computed from a plane wave source for all lattices.  \label{ghost}} 
\end{figure}

In figure \ref{ghost} (left) it is shown the
ghost dressing function $Z_{ghost}( q^2 )=q^2 G(q^2)$ for the 
 $16^3\times128$ lattice, gauge fixed using CEASD method \cite{ceasd},
 computed with both methods. As in the gluon case, we can see 
differences, in the infrared, between pure 
temporal and pure spatial data. These differences 
vanish for sufficiently high momenta.

\vspace*{0.4cm}

On the right hand side of figure \ref{ghost}, 
one can see the ghost dressing function only for 
the plane-wave data, for the available lattices. As in the gluon case 
\cite{brasil}, we are able to evaluate the effect of Gribov 
copies on the lattice $16^3\times128$ by considering different gauge fixing 
methods. The data show clear effects of Gribov copies over a 
large range of momenta, as expected from other studies \cite{cucc97,stern05}. 
Also, we see finite volume effects 
if one compares propagators from lattices with different 
spatial sizes. 

In what concerns the infrared region, we were unable to fit a pure power 
law, even considering polinomial corrections \cite{madrid}.

\section{Running coupling constant}

From the gluon and ghost dressing functions, one can define a 
running coupling constant --- see eq. \ref{alfaS}.

The DSE infrared analysis predicts a running coupling constant at zero momentum
different from zero, $\alpha_{S}(0)=2.972$ \cite{lerche}. On the other 
hand, the DSE solution on a torus \cite{dsetorus}, and results from 
lattice simulations \cite{stern05, furui2, madrid}, show a 
decreasing coupling constant 
for small momenta. Using an asymmetric lattice allow us to study smaller 
momenta having in mind to provide, at least, a hint to this puzzle.

\begin{figure}[!t]
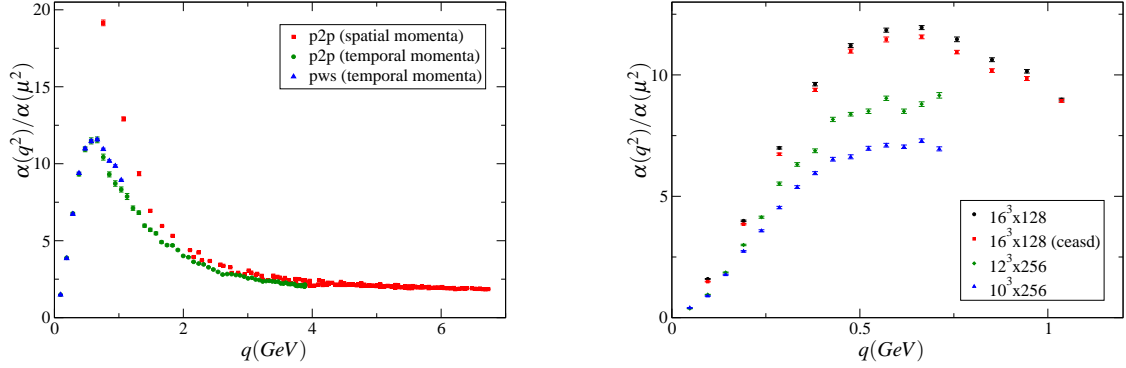

\psfrag{EIXOX}{{\scriptsize $q(GeV)$}}
\psfrag{EIXOY}{{\scriptsize $\alpha(q^2)/\alpha(\mu^2)$}}
   \begin{minipage}[b]{0.45\textwidth}
   \centering
   \includegraphics[width=6.5cm]{alpha.16.3.128.ceasd.eps}
  \end{minipage} 
\hspace{0.08\textwidth}
  \begin{minipage}[b]{0.45\textwidth}
  \centering
  \includegraphics[width=6.5cm]{alpha.cucc.eps}
  \end{minipage} 
\caption{ On the left, the running coupling constant for the 
$16^3\times128$ lattice, gauge fixed with CEASD method. 
 On the right, the running coupling constant for all lattices. The data were 
computed using a plane wave method.  \label{alpha}} 
\end{figure}

Again, our lattice data show finite volume effects, if one compares pure 
temporal and pure spatial momenta, see figure \ref{alpha} (left).
Comparing the results for all available lattices (plane-wave source), 
see figure \ref{alpha} (right), we can see, as in the ghost case, 
finite volume effects, and clear Gribov copies effects.

In what concerns the infrared behaviour, we tried to fit the lowest momenta 
to a pure power law, $(q^2)^{\kappa_{\hspace{0.5mm}\alpha}}$. We concluded that this 
power law is only compatible with the data from $16^3\times128$ lattice, 
gauge fixed with CEASD method, giving $\kappa_{\alpha}\sim 0.688$, with 
$\chi^{2}/d.o.f.\sim 0.011$. The reader should be aware that it is also 
possible, in some cases, to fit the infrared data to 
$\alpha(0) (1+ aq^2+\ldots)$ and get 
a $\alpha(0)\neq 0$. Therefore, we can not give a definitive answer about the 
behaviour of the running coupling constant for $q=0$. Note, however, that 
$\alpha_S (q^2)$ for the smallest momenta, seems to increase as 
a function of the volume.

\section{Future work}

Currently, we are engaged in improving the statistics for our larger 
lattices and the extrapolations to the infinite volume limit for the 
gluon and ghost propagators. Furthermore, we plan to perform simulations 
with larger lattices and, hopefully, combine all our results to provide 
a reliable answer on the behaviour of the infrared QCD Green's functions.

\section*{Acknowledgements}
 
The speaker would like to thank Funda\c c\~ao Luso-Americana para 
o Desenvolvimento (FLAD), Funda\c c\~ao Calouste Gulbenkian and 
University of Arizona Foundation to make this interesting conference 
possible to him. This work was supported by FCT via grant 
SFRH/BD/10740/2002, and project POCI/FP/63436/2005.

\end{document}